\documentclass[12pt]{article}
\usepackage{lathuile2}
\usepackage[latin1]{inputenc}
\usepackage{graphicx}
\usepackage{overcite}
\usepackage{mcite}
\usepackage{xspace}
\usepackage{amssymb}
\newcommand{\alphas}{\ensuremath{\alpha_{S}}}
\newcommand{\Ftwo}{\ensuremath{F_{2}}}
\newcommand{\Fthree}{\ensuremath{F_{3}}}
\newcommand{\Fl}{\ensuremath{F_{L}}}
\newcommand{\Qsq}{\ensuremath{Q^{2}}}
\newcommand{\Et}{\ensuremath{E_{T}}}
\newcommand{\Etjet}{\ensuremath{E_{T,\mathrm{jet}}}}

\newcommand{\GeV}{\ensuremath{\mathrm{Ge\kern -0.1em V}}}
\newcommand{\xGeV}{\ensuremath{\:\GeV}}

\newcommand{\xipb}{\ensuremath{\:\mathrm{pb}^{-1}}}

\newcommand{\xifb}{\ensuremath{\:\mathrm{fb}^{-1}}}
\newcommand{\mZ}{\ensuremath{m_{Z}}}
\graphicspath{%
  {figs/arxiv/}%
  }
\DeclareGraphicsExtensions{.eps,.ps}
\newlength{\figwidth}
\begin{document}
\title{%
{\normalsize\normalfont \hfill BONN-HE-2002-02\\[-0.1em]
\hfill Sepember 2002}\\[2em]
QCD STUDIES AT HERA}
\author{
Ian C.\ Brock\\
{\em Physikalisches Institut der Universität Bonn} \\
Representing the ZEUS and H1 Collaborations        \\
}
\maketitle
\thispagestyle{empty}

%
%
\baselineskip=14.5pt
\begin{abstract}
I present a small selection of the many QCD studies that have been performed at
HERA. I concentrate on results that have been made available in the past
year. Results are presented on the QCD description of the parton density
functions, determinations of \alphas, jets and heavy flavour production. Where
possible comparisons are made with next to leading order (NLO) QCD
predictions. The neutral and charged current cross sections at very high \Qsq{}
are also discussed.
\end{abstract}
\baselineskip=17pt
\newpage
%
%
\section{Introduction}

It is fair to say that most of the HERA I programme has had something to do
with QCD. Such a talk could, and maybe should, cover about 50\% of HERA
physics. As this is impossible in the time available I will concentrate on a
few selected topics. With the data taken up to the end of 2000, each of the
collider experiments H1 and ZEUS have over $120\xipb$ of data available for
analysis. Considerable progress on reducing systematic uncertainties has also
been made, leading to a number of results that have accuracies at the few
percent level. At the same time next to leading order (NLO) QCD calculations
are available for many of the processes considered. In some cases these
predictions also have the same level of accuracy as the data. Given the large
corrections between leading order (LO) and NLO calculations it is also clear
that NNLO is needed in many places. Some first calculations have recently been
released and the experimental groups are eagerly waiting for more. For a
number of the measurements, particularly of \alphas{}, the dominant systematic
uncertainty is due to the (somewhat) arbitrary definition of the uncertainty
on the factorisation and renormalisation scales. More theoretical input on
reasonable values to take for these uncertainties would be welcome.

In this talk I will discuss:
Parton Density Functions (PDFs);
the determination of \alphas;
three jet production;
dijets in photoproduction;
measurements at the highest values of \Qsq{} and
heavy flavour production.

Interactions at HERA are characterised by a number of interdependent kinematic
variables:
\begin{itemize}
\setlength{\parskip}{0pt}\setlength{\itemsep}{0pt}
\item $s$, the centre-of-mass energy squared,
\item $Q^{2} = -q^{2}$, the negative four-momentum transfer squared,
\item $x$, the Bjorken-$x$ variable,
\item $y$ the inelasticity, and
\item $W$ the photon-proton centre-of-mass energy.
\end{itemize}
The first 4 variables are related by the equation:
\begin{equation}
  \label{eq:qsxy}
  Q^{2} = s x y
\end{equation}
The centre-of-mass energy was 300\xGeV{} until 1997. In 1998 the proton beam
energy  was increased from 820\xGeV{} to 920\xGeV, resulting in a
centre-of-mass energy of 318\xGeV.

%
%
\section{Parton Density Functions}
\label{sec:pdf}

The determination of the parton density functions (PDFs) was, and still is,
one of the main goals of the HERA physics programme.  The QCD factorisation
theorem allows one to separate the cross-section into a short-range part that
is described by QCD and a long-range part.  One can use the so-called
DGLAP~\cite{jetp:46:641,*np:b126:298,*sovjnp:15:438,*sovjnp:20:94} equations
to evolve the long-range part from a given value to \Qsq{} to higher values.
The experimental data have to constrain the PDFs at a starting value
$Q_{0}^{2}$.

\begin{figure}[htbp]
  \begin{center}
    \includegraphics[width=\figwidth, bb=30 25 528 763, clip=]{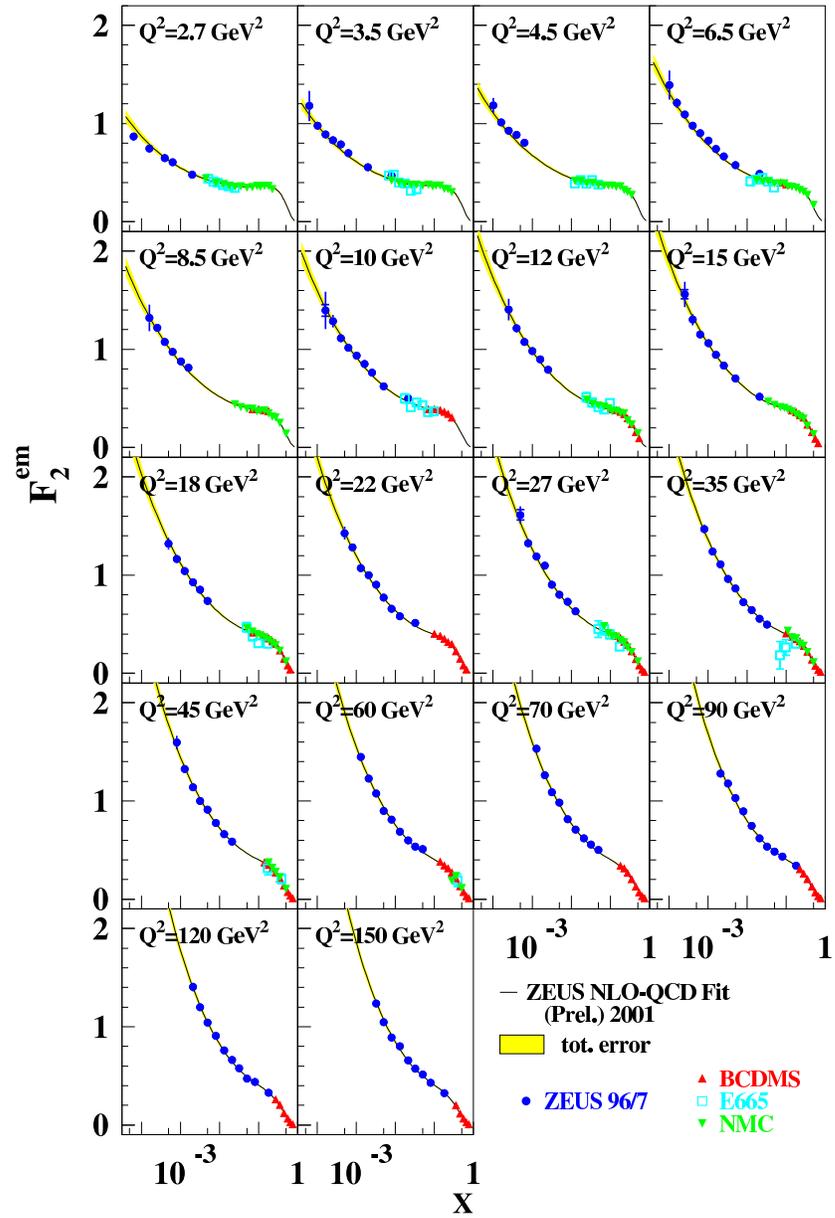}
    \caption{\textit{%
        ZEUS Measurements of the structure function $F_{2}$
        }
      }
    \label{fig:zeus-f2}
  \end{center}
\end{figure}

The neutral current cross section for $e^{\pm} p$ scattering can be written in
the form:
\begin{equation}
  \label{eq:ncxsect}
  \frac{d^{2}\sigma}{dx\,dQ^{2}} = \frac{2 \pi \alpha^{2}}{x Q^{4}}
  \left( Y_{+} \Ftwo - y^{2} \Fl \mp Y_{-} x F_{3} \right)
\end{equation}
where $Y_{\pm} = \left( 1 \pm (1-y)^{2} \right)$.  At small $y$ and for $\Qsq
\ll \mZ^{2}$ the cross section is dominated by \Ftwo. An example of the
determination of the structure function $F_{2}$ by ZEUS~\cite{misc:eps01:628}
is shown in fig.~\ref{fig:zeus-f2}. Many systematic effects have to be studied
in such measurements; therefore the data shown here are only those taken in
1996 and 1997. In the kinematic range shown the errors are dominated by
systematics.  These are now at the level of about 2\% over much of the
kinematic range. The figure also includes some fixed target measurements. In
the regions where the kinematic range overlaps, good agreement with previous
measurements is seen. The HERA data extends the measurements to much lower
values of $x$ and the rapid rise of the structure function there is clearly
visible and well measured.

\begin{figure}[htbp]
  \begin{center}
    \includegraphics[width=\figwidth, bb=30 145 565 645, clip=]{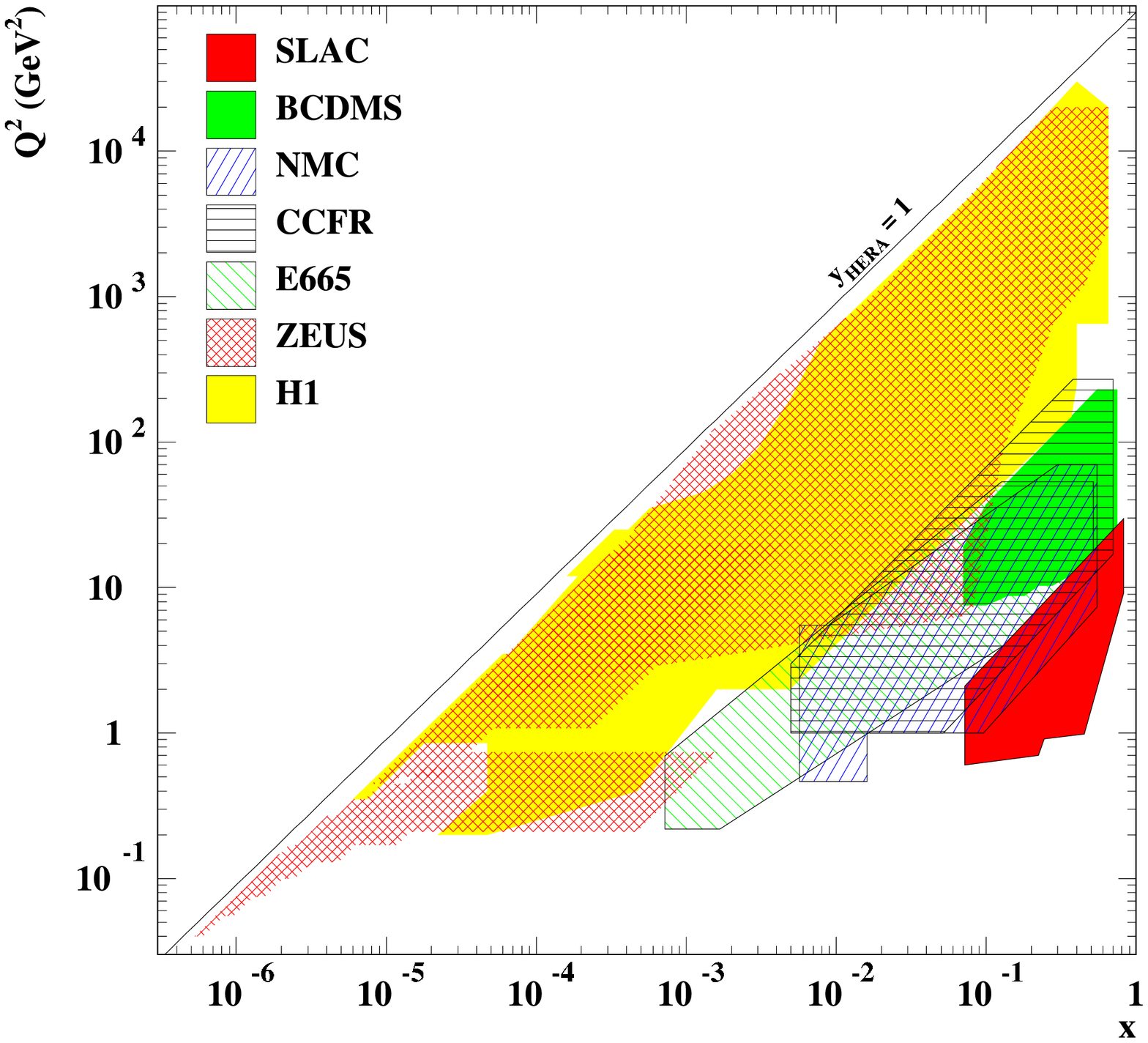}
    \caption{\textit{%
        The kinematic range covered by fixed target and the HERA
        experiments. 
        }
      }
    \label{fig:h1-kine}
  \end{center}
\end{figure}

The complete set of ZEUS and H1 deep inelastic scattering (DIS) data cover a
very wide range in $x$ and \Qsq. This is illustrated in
fig.~\ref{fig:h1-kine}.
At present it is still
necessary to also include some fixed target data in order to fix the PDFs at
large $x$. With the full HERA~I dataset this may no longer be necessary. Both
ZEUS and H1 fit their data in a similar way. However, some of the details are
different, which can affect the extracted parton densities. As a first
step the PDFs have to be parametrised. The form used is:
\begin{equation}
  \label{eq:pdf}
  f = p_{1} x^{p_{2}} ( 1 - x )^{p_{3}} \left(
    1 + p_{4} \sqrt{x} + p_{5} x 
  \right)
\end{equation}
where $p_{1}, p_{2}, p_{3}, p_{4}$ and $p_{5}$ are free parameters.
Both collaborations use the NLO DGLAP equations. The differences between the
inputs are summarised in Table~\ref{tab:pdf-fit}.
\begin{table}
  \centering
  \caption{\textit{%
      Differences between H1 and ZEUS fits of PDFs.
      }
    }
  \vskip 0.1 in
  \begin{tabular}{|l|p{5.0cm}|p{5.0cm}|}
    \hline
    & \multicolumn{1}{c|}{ZEUS~\cite{misc:eps01:628}} & 
    \multicolumn{1}{c|}{H1~\cite{h1:eps01:787}} \\ \hline\hline
    Fixed target data & BCDMS, E665, NMC & BCDMS \\
    $Q_{0}^{2}$ &  $7\xGeV^{2}$ &  $4\xGeV^{2}$\\
    $p_{4}$ & Set to zero plus other constraints & Zero for gluon \\
    Heavy quarks & RT variable flavour number scheme & Fixed flavour number
    scheme \\\hline
  \end{tabular}
  \label{tab:pdf-fit}
\end{table}

\begin{figure}[htbp]
  \begin{center}
    \includegraphics[width=\figwidth, clip=]{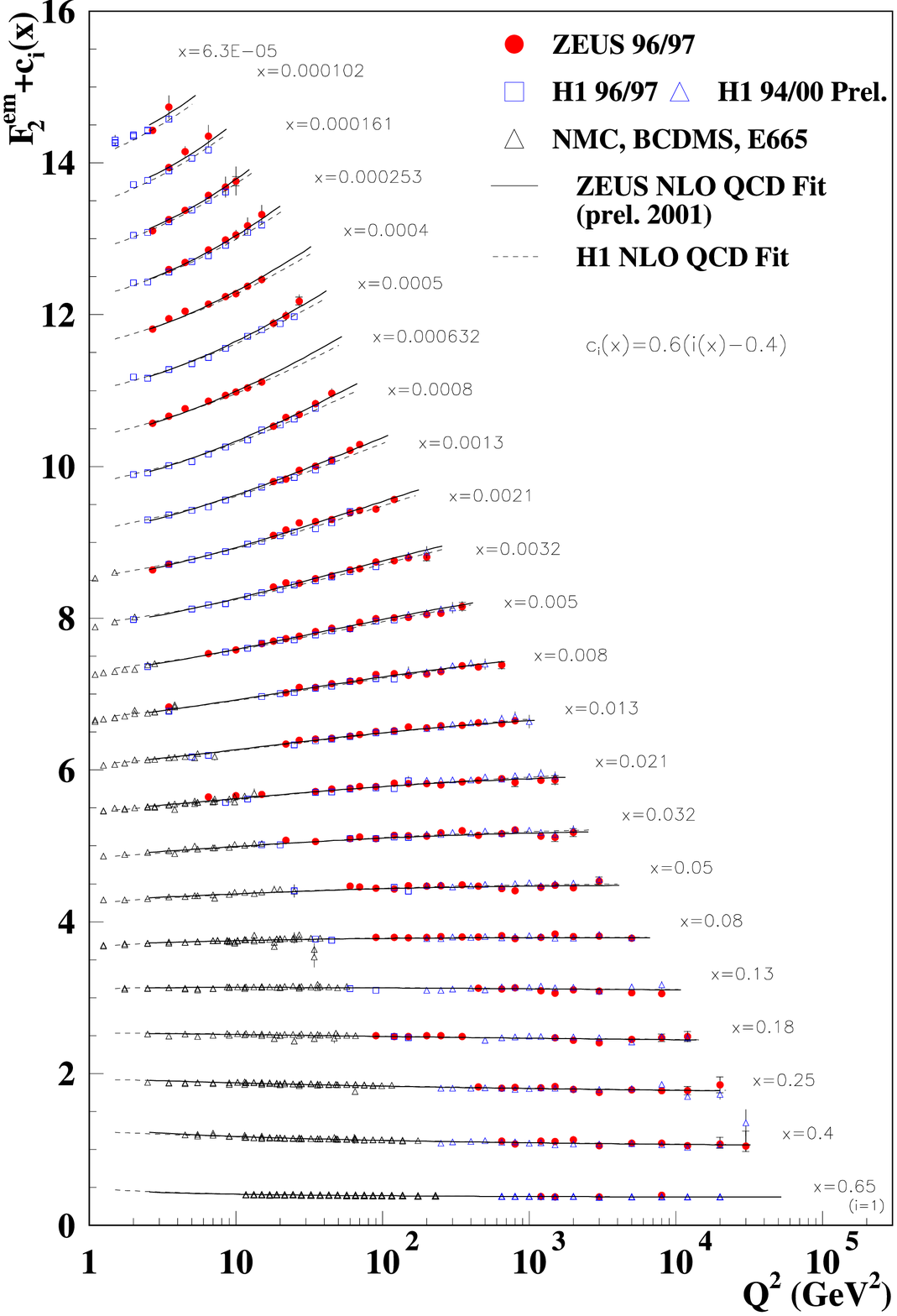}
    \caption{\textit{%
        Results of NLO QCD fits to HERA data.
        }
      }
    \label{fig:pdf-fit}
  \end{center}
\end{figure}

\begin{figure}[htbp]
  \begin{center}
    \begin{tabular}{ll}
      \includegraphics[width=7cm, bb=20 145 520 665, clip=]{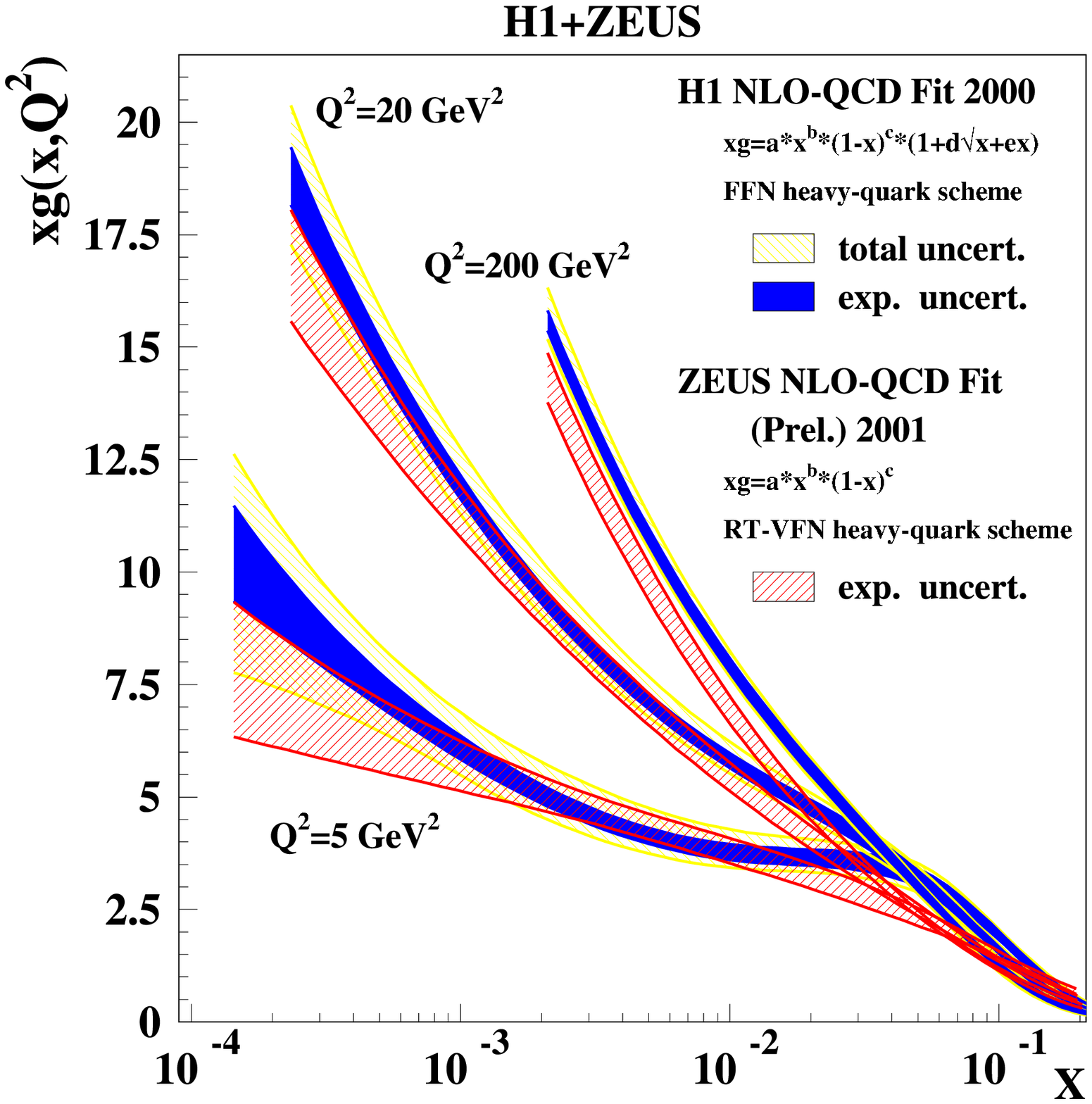} &
      \includegraphics[width=7cm, bb=35 145 525 675, clip=]{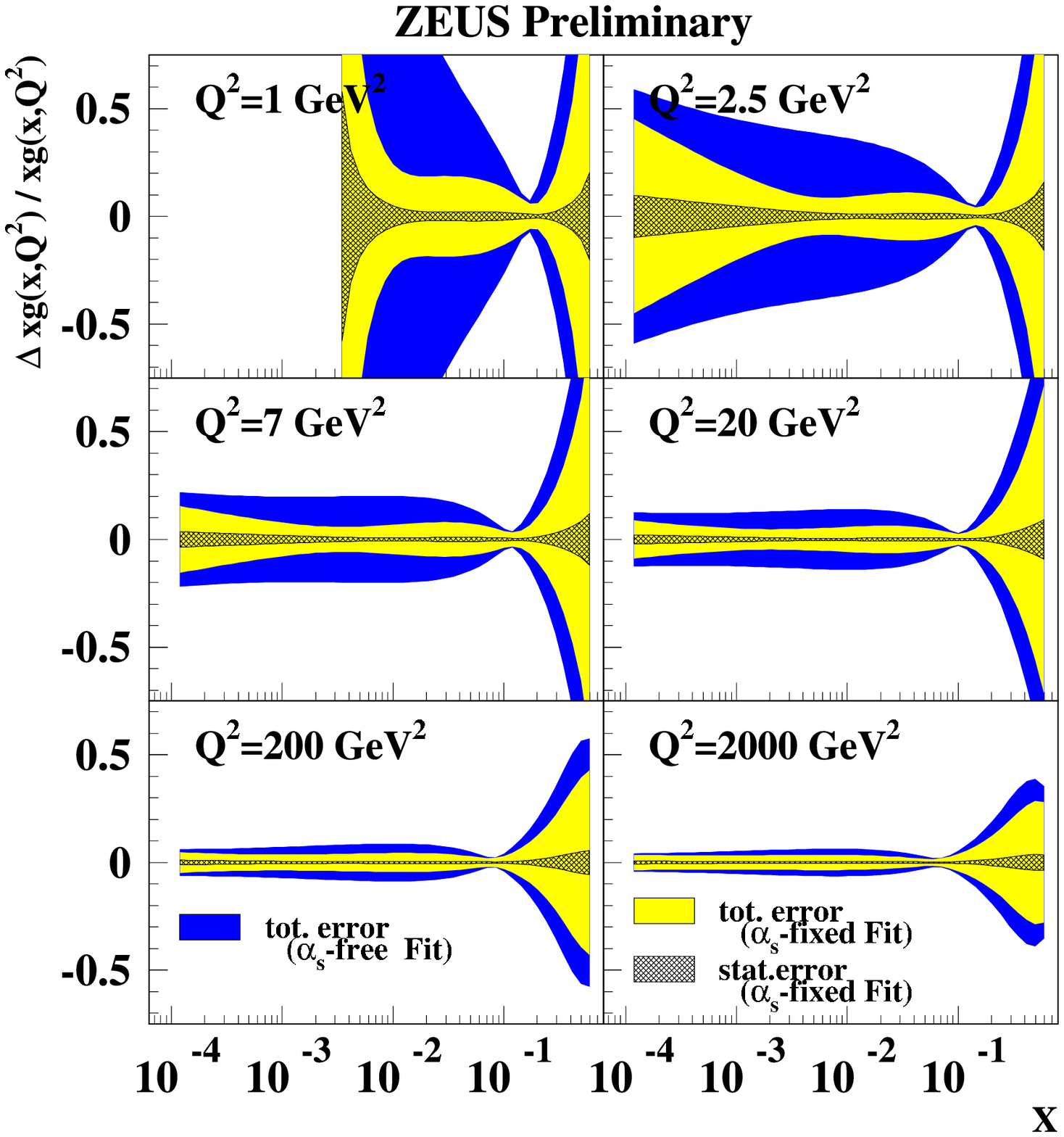}
    \end{tabular}
    \caption{\textit{%
        The left-hand plot show the gluon densities from ZEUS and H1 at $\Qsq
        = 20\xGeV^{2}$. The right-hand plot shows the ZEUS error on 
        $x g(x)$ for different \Qsq{} values as a function of $x$.
        }
      }
    \label{fig:gluon}
  \end{center}
\end{figure}

Examples of the fit results are shown in fig.~\ref{fig:pdf-fit}.
As can be seen the ZEUS and H1 data and their fits agree over a very wide
range in the kinematic plane. There are some small differences at small $x$,
which I will briefly discuss below. At high \Qsq{} the differences are due to
the limited statistics. In a new development, the fits now also calculate
errors on the PDFs. The correlations between the experimental errors are taken
into account. The gluon distribution that is extracted by the two
collaborations agrees in general well, but there are again some differences.
This can be seen in fig.~\ref{fig:gluon}.
The evolution of the gluon density as a function of \Qsq{} is clearly visible.
The differences between the fits at low $x$ and around $x \approx 0.05$ are
being discussed by the collaborations and are probably due to a combination of
the heavy flavour scheme and the parameterisation used. If one looks at the
right-hand figure, the strong correlation of the error on $x g(x)$ with
\alphas{} is seen. One also sees though that for high \Qsq{} and certain
ranges of $x$ the effect is smaller.

\begin{figure}[htbp]
  \begin{center}
    \begin{tabular}{ll}
      \includegraphics[width=7cm, bb=0 0 540 450, clip=]{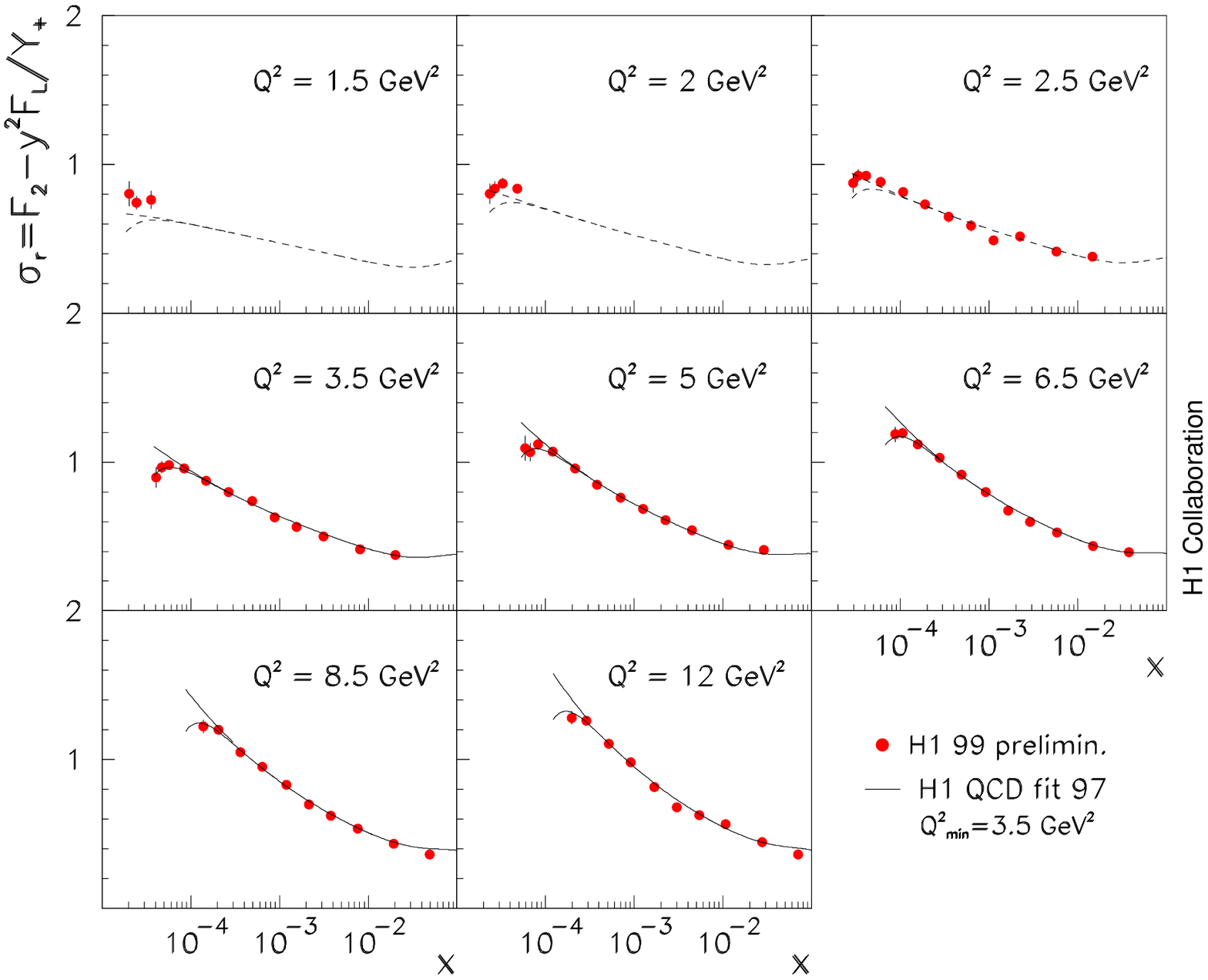} &
      \includegraphics[width=7cm, clip=]{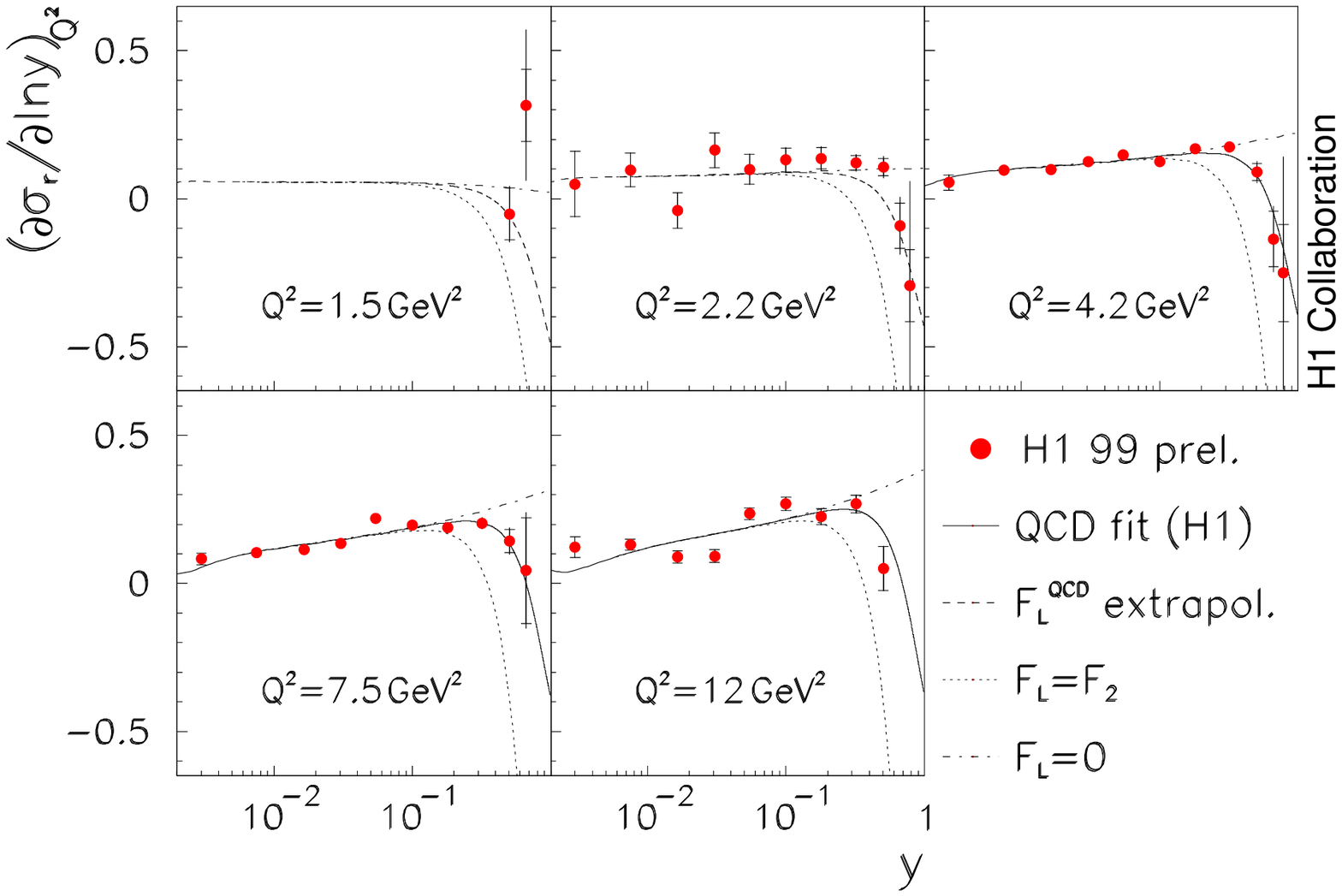}
    \end{tabular}
    \caption{\textit{%
        Two methods used by H1 to extract \Fl.
        }
      }
    \label{fig:fl-method}
  \end{center}
\end{figure}

\begin{figure}[htbp]
  \begin{center}
    \includegraphics[width=8cm, clip=]{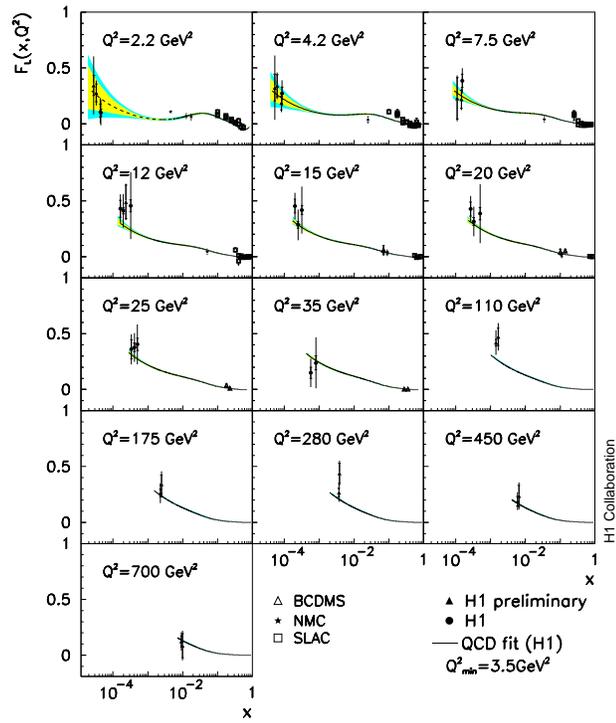}
    \caption{\textit{%
        \Fl{} as measured by H1 (low $x$) and lepton-nucleon fixed-target
        experiments (high $x$).
        }
      }
    \label{fig:fl-result}
  \end{center}
\end{figure}

The reduced cross section is defined as:
\begin{equation}
  \label{eq:xsec-red}
  \tilde{\sigma} = \frac{Q^{4} x}{2 \pi \alpha^{2} Y_{+}}
  \frac{d^{2} \sigma}{dx\,d Q^{2}}
\end{equation}
and can be written as:
\begin{equation}
  \label{eq:fldef}
  \tilde{\sigma} = F_{2} - \frac{y^{2}}{Y_{+}} F_{L}
\end{equation}
where $F_{L}$ is the longitudinal structure function and $x\Fthree$ has been
neglected. The best way to measure $F_{L}$ is to take data at significantly
different centre-of-mass energies. As the centre-of-mass energy of HERA has
not changed substantially the collaborations have tried other methods. ZEUS
used events with significant initial state
radiation~\cite{thesis:bornheim:1999}. H1 has developed a new
method~\cite{epj:c21:33}. They measure \Ftwo{} in a region where the \Fl{}
contribution is small and use the DGLAP equations to extrapolate this
measurement to regions where the contribution of \Fl{} to the reduced cross
section is expected to be significant (high $y$). By comparing the reduced
cross section to the extrapolation they can extract \Fl{} (see
fig.~\ref{fig:fl-method}). 
This method works best at high \Qsq. At lower
\Qsq{} they look at $\partial\sigma / \partial \ln y$. 
As \Fl{} depends directly on the gluon density, the agreement between \Fl{}
and the QCD fit for both methods confirms the validity of the fit.
The results are shown in fig.~\ref{fig:fl-result}.

%
%
\section{Determinations of \alphas}
\label{sec:alphas}

A number of new methods have been developed that allow a precise
determination of \alphas. Most of them use jets reconstructed in either the
laboratory or the Breit frame.

Jet are usually found using clustering algorithms. H1 recently investigated
how low they could set the cut that defines a jet and still find agreement
with perturbative QCD~\cite{epj:c24:33}. They found that even with a cut at
$y_{\mathrm{cut}} \approx 0.001$, where about 30\% of events in DIS could be
classified as having two or more jets, good agreement is still seen.

Many of the jet studies are made in the Breit frame. In this frame jets with a
significant transverse energy are produced via the QCD-Compton or the
boson-gluon-fusion process in LO. Their production rate depends on \alphas,
but the data is in general more sensitive to $\alphas \cdot x g(x)$.

\begin{figure}[htbp]
  \begin{center}
    \begin{tabular}{ll}
      \includegraphics[width=6cm, bb=130 175 505 805, clip=]{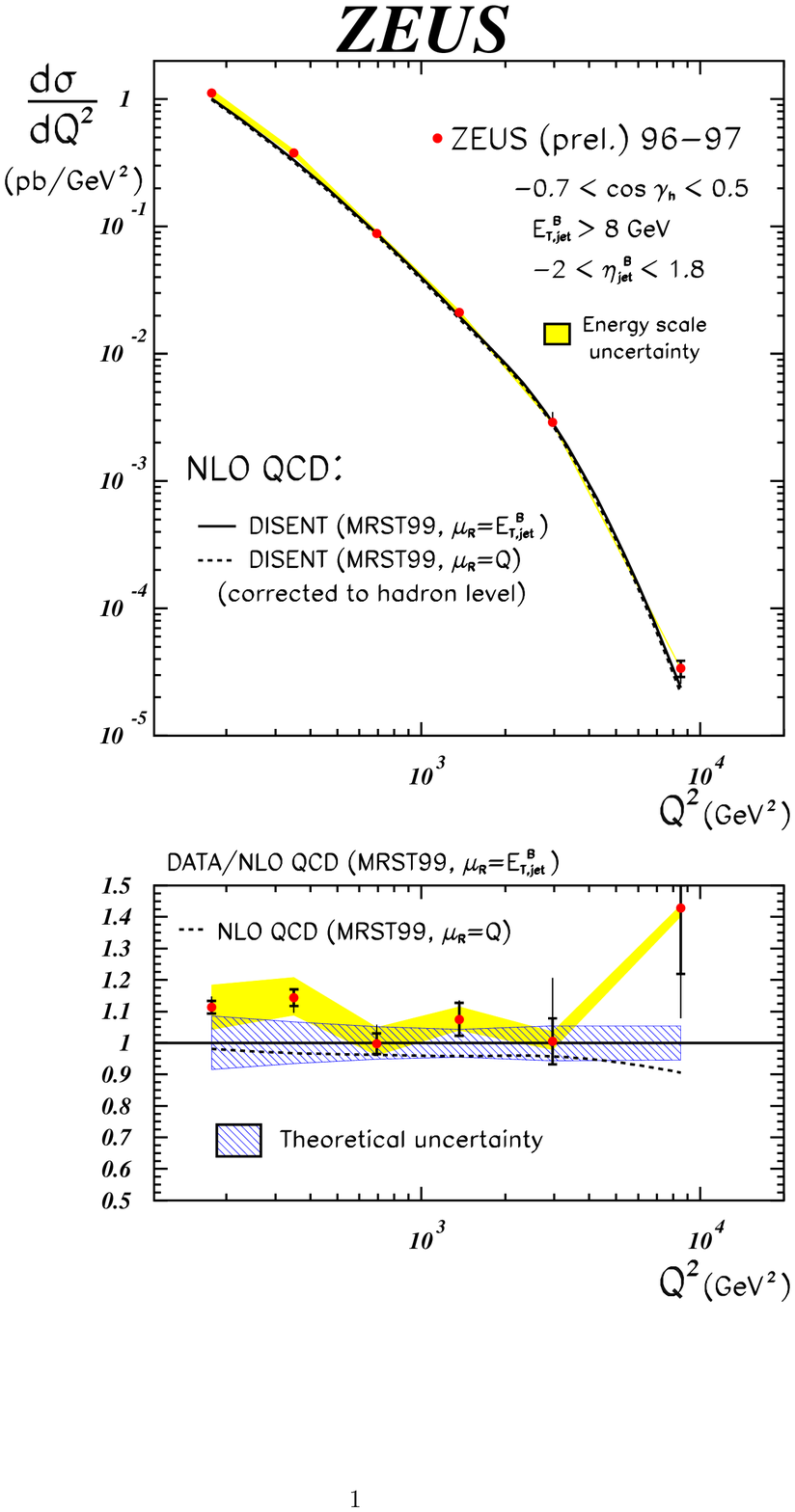} &
      \includegraphics[width=6cm, bb=130 175 505 805, clip=]{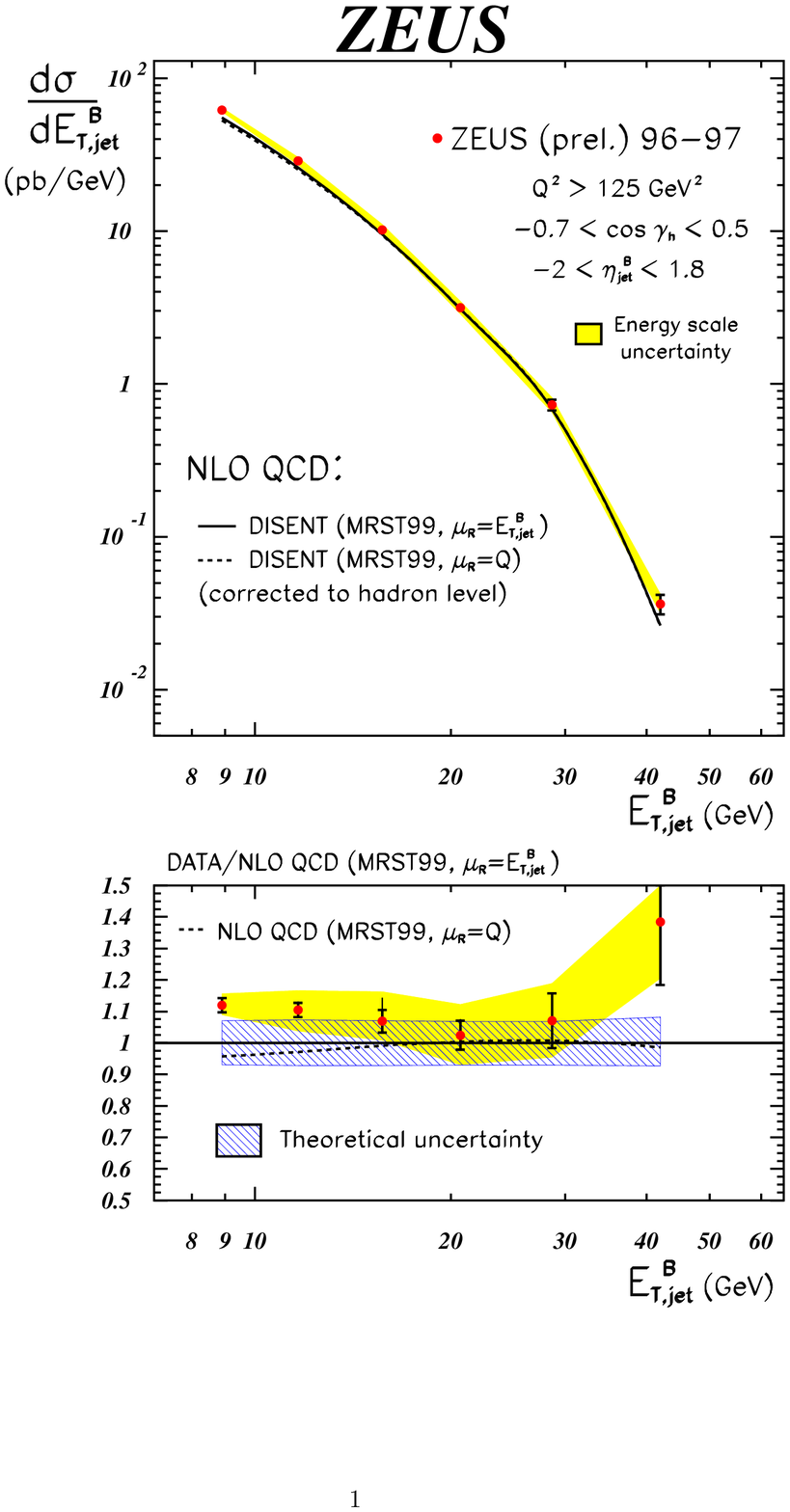}
    \end{tabular}
    \caption{\textit{%
        Jet cross sections as a function of \Qsq{} and \Et.
        }
      }
    \label{fig:onejet}
  \end{center}
\end{figure}

\begin{figure}[htbp]
  \begin{center}
    \includegraphics[width=7cm, clip=]{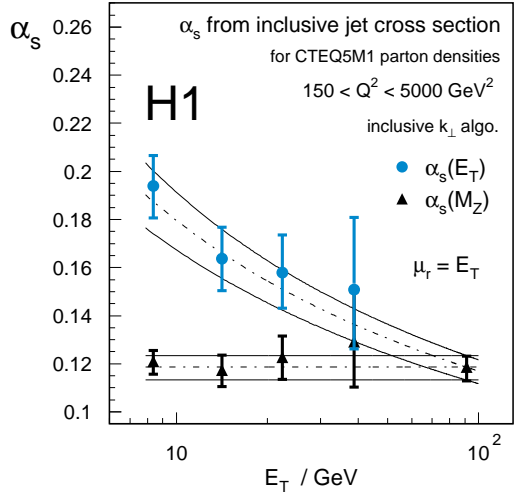}
    \caption{\textit{%
        \alphas{} measured by H1 using the inclusive jet cross section
        as a function of \Et.
        }
      }
    \label{fig:alphas-et}
  \end{center}
\end{figure}

The single jet cross sections measured by ZEUS~\cite{zeus:eps01:637} as a
function of \Qsq{} and \Et{} are shown in fig.~\ref{fig:onejet}.
Good agreement is seen as a function of both variables over a very wide range.
The main theoretical error comes from the uncertainty on the renormalisation
and factorisation scales. As indicated in the introduction, this is a common
problem for many such jet analyses. \alphas{} has been extracted from these
measurements. H1 has also
extracted \alphas{} from the inclusive jet cross section~\cite{epj:c19:289}.
By studying the size of the correction factors from the observed jets to the
parton level it appears that such methods are reliable for $\Qsq >
150\xGeV^{2}$ or $\Et > 15\xGeV$.  The value of \alphas{} as a function of
\Et{} is shown in fig.~\ref{fig:alphas-et}. The running of \alphas{} is
clearly seen. Correcting each value to \mZ{} the measurements are all
consistent.  The theory error is at the level of about 5\%.

\begin{figure}[htbp]
  \begin{center}
    \includegraphics[width=8.0cm, clip=]{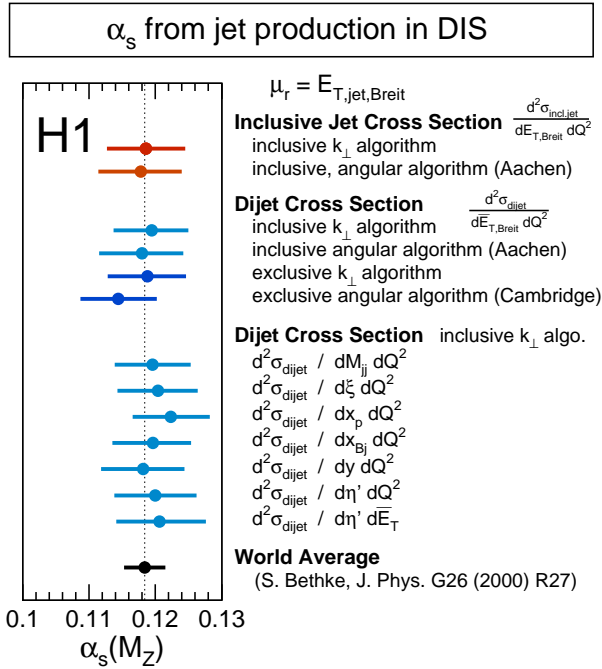}
    \caption{\textit{%
        Comparison of \alphas{} results from fits to different 
        double-differential jet distributions.
        }
      }
    \label{fig:h1-dijet}
  \end{center}
\end{figure}

The dijet cross sections can also be used to determine \alphas. H1 studied the
cross section as a function of many kinematic variables and for different jet
algorithms~\cite{epj:c19:289}.  They find very good agreement between the
different methods as can be seen in fig.~\ref{fig:h1-dijet}.

\begin{figure}[htbp]
  \begin{center}
    \includegraphics[width=9.0cm, clip=]{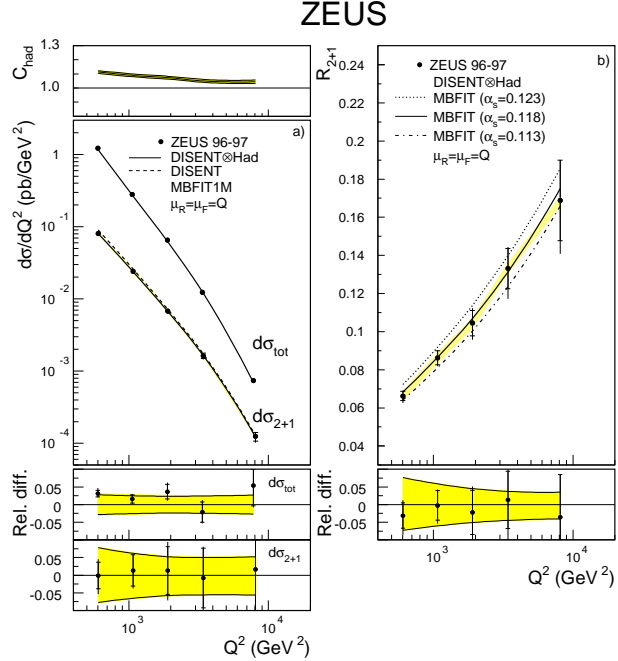}
    \caption{\textit{%
        The single and dijet cross sections as a function of \Qsq. The size of
        the hadron to parton correction in the case of dijets 
        is indicated at the top of the left-hand figure. 
        The right-hand figure shows the ratio of the dijet to the single jet
        cross section and compares it with the predictions for different
        values of \alphas.
        }
      }
    \label{fig:zeus-dijet}
  \end{center}
\end{figure}

ZEUS has compared the cross sections for single and dijet
production~\cite{pl:b507:70}.  
The cross sections and their ratio are shown in fig.~\ref{fig:zeus-dijet}.
The ratio of the cross sections is proportional
to \alphas. This can be seen in the right-hand figure, where the ratio is
compared to curves for different values of \alphas. H1 has studied the
dependence of the dijet cross section on the jet algorithm at high \Qsq. By
comparing the LO and NLO predictions, it is clear that NLO is needed. They
extract \alphas{} in the range $150 \leq \Qsq \leq 5000\xGeV^{2}$ so that the
correction factors from LO to NLO are not too large. 

\begin{figure}[htbp]
  \begin{center}
    \includegraphics[width=7cm, bb=145 400 510 800, clip=]{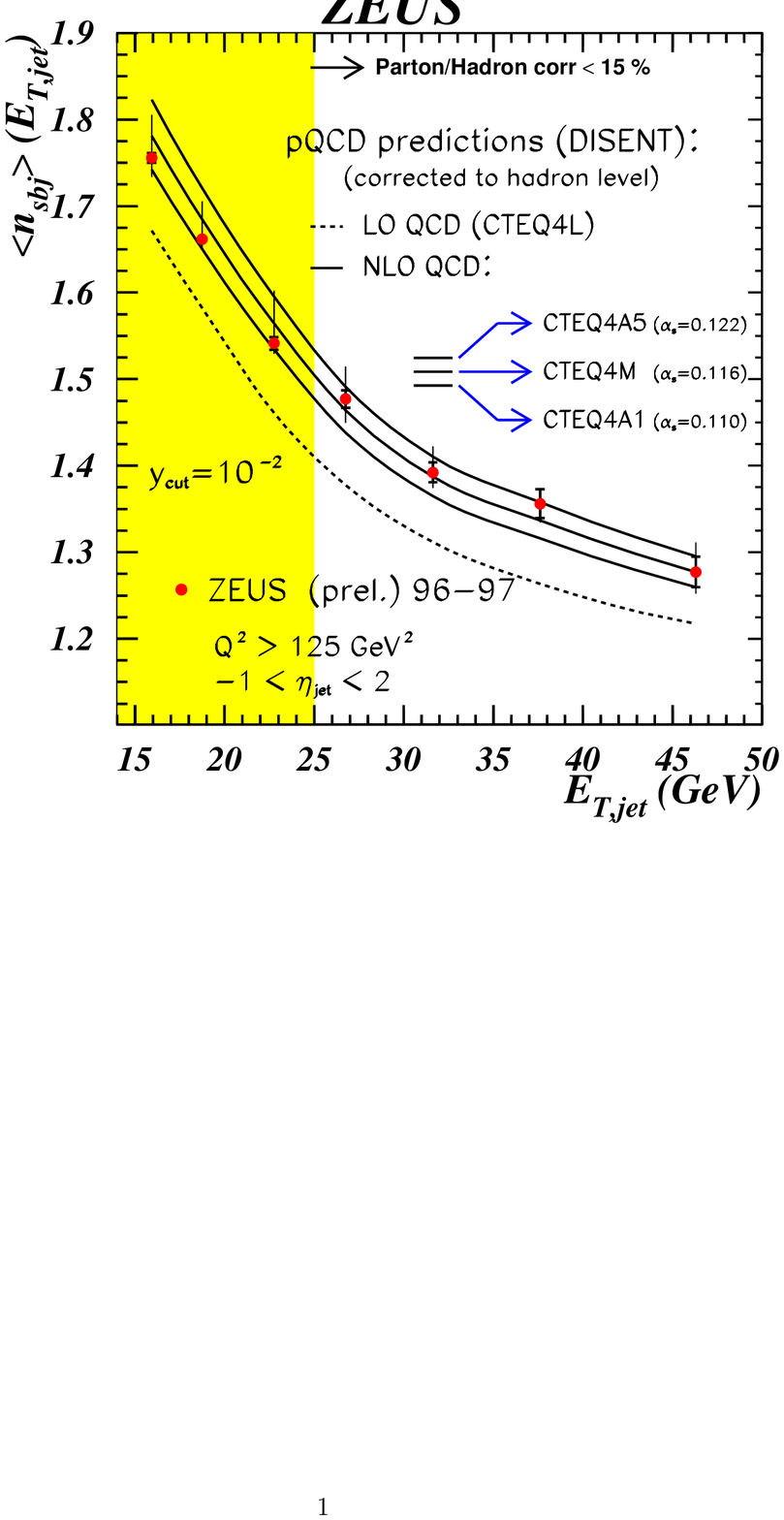}
    \caption{\textit{%
        Number of subjets as a function of \Etjet.
        }
      }
    \label{fig:subjet}
  \end{center}
\end{figure}

Jet substructure can even be used to extract \alphas. ZEUS look for jet-like
components inside a jet~\cite{zeus:eps01:641}. The jets are reconstructed in
the lab system, in order to keep the large single-jet sample. Once again a
region has to be found where the parton to hadron corrections are relatively
small, defined in this analysis to be $< 15\%$. The number of subjets is then
measured in the data and compared to NLO predictions. The measurements are
sensitive to \alphas, as can be seen in fig.~\ref{fig:subjet}.

\begin{figure}[htbp]
  \begin{center}
    \includegraphics[width=8cm, bb=70 35 405 540, clip=]{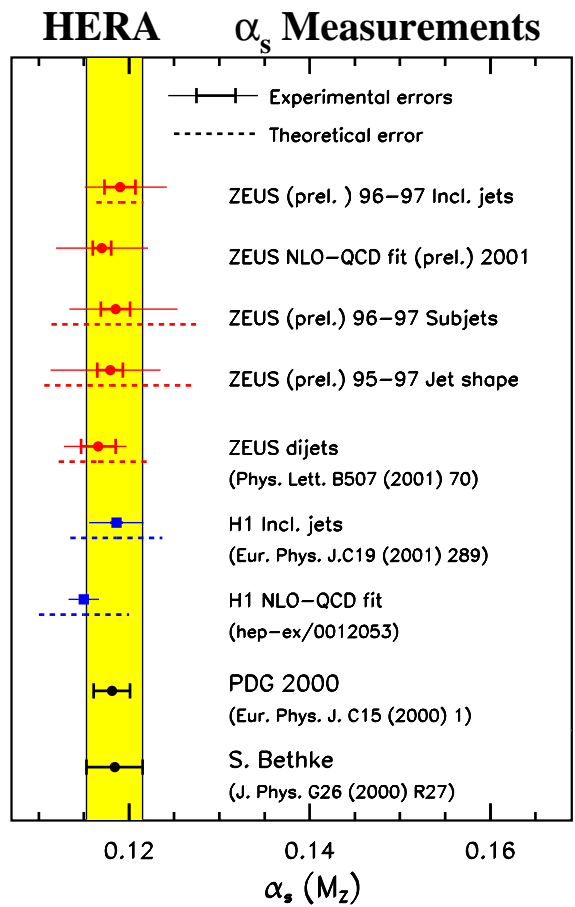}
    \caption{\textit{%
        Different determinations of \alphas{} at HERA.
        }
      }
    \label{fig:alphas-sum}
  \end{center}
\end{figure}

The various HERA \alphas{} measurements are summarised in
fig.~\ref{fig:alphas-sum}. 
A number of the measurements are at the same level of statistical accuracy as
the world average. However, their influence on the average depends on
being able to reduce the theoretical uncertainties, particularly those
associated with the renormalisation and factorisation scales.

%
%
\section{Three Jet Production}
\label{sec:3jet}

\begin{figure}[htbp]
  \begin{center}
    \begin{tabular}{ll}
      \includegraphics[width=7cm, clip=]{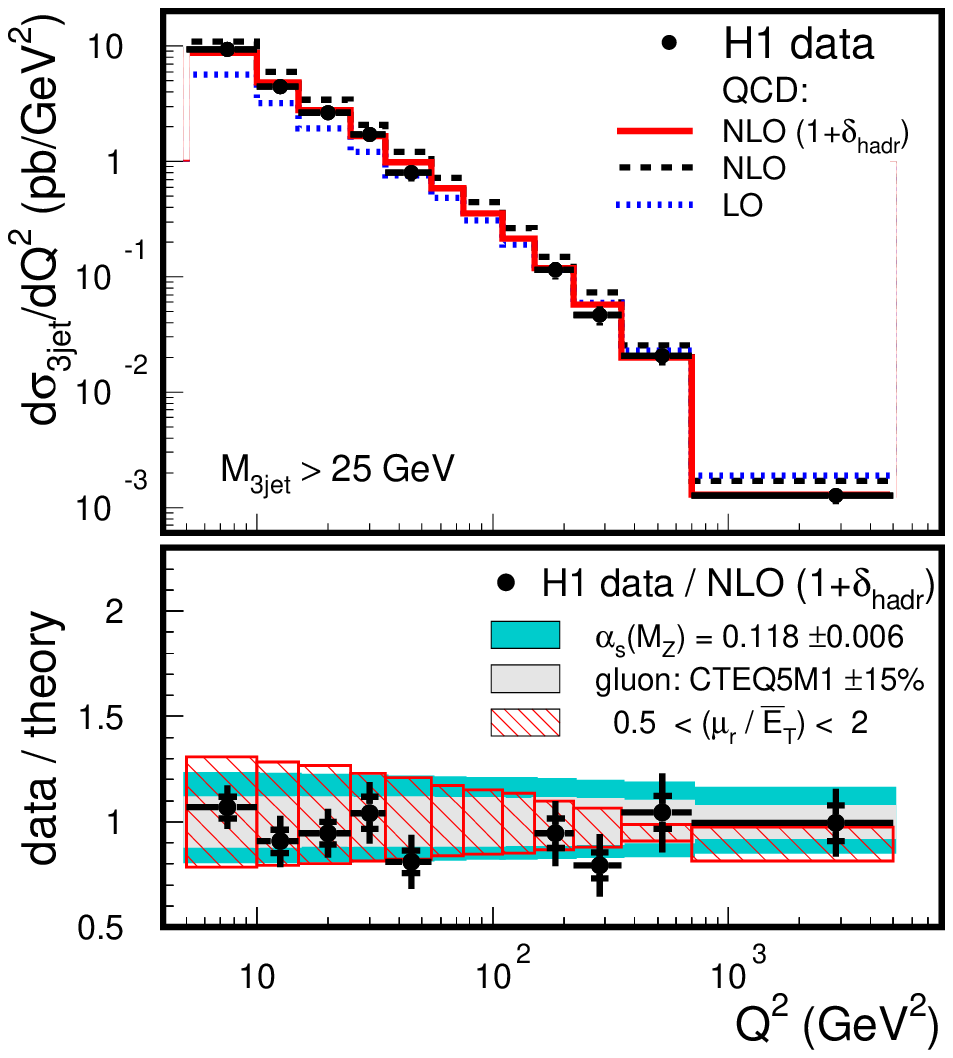} &
      \includegraphics[width=7cm, clip=]{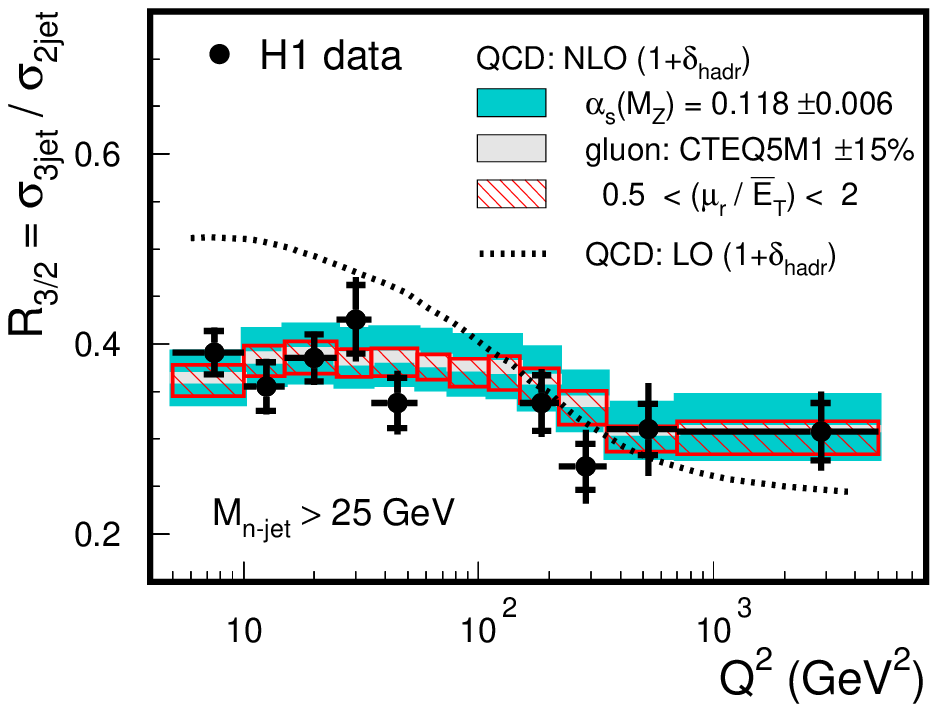}
    \end{tabular}
    \caption{\textit{%
        The three jet cross section as a function of \Qsq. The right-hand plot
        shows the ratio of three-jet to two-jet production.
        }
      }
    \label{fig:threejet}
  \end{center}
\end{figure}

Moving up a further power of \alphas, to $\alphas^{3}$, H1 has looked at three
jet production~\cite{pl:b515:17}. The cross section is proportional to
$\alphas^{2}$ at lowest order and has been compared with NLO calculations
(order $\alphas^{3}$) as a function of \Qsq. The results are shown in
fig.~\ref{fig:threejet}.  The NLO contributions are clearly necessary,
particularly at low \Qsq. With the contributions included, good agreement
between data and theory is seen over the whole range.

%
%
\section{Dijets in Photoproduction}
\label{sec:2jet}

\begin{figure}[htbp]
  \begin{center}
    \begin{tabular}{ll}
      \includegraphics[width=\figwidth, clip=]{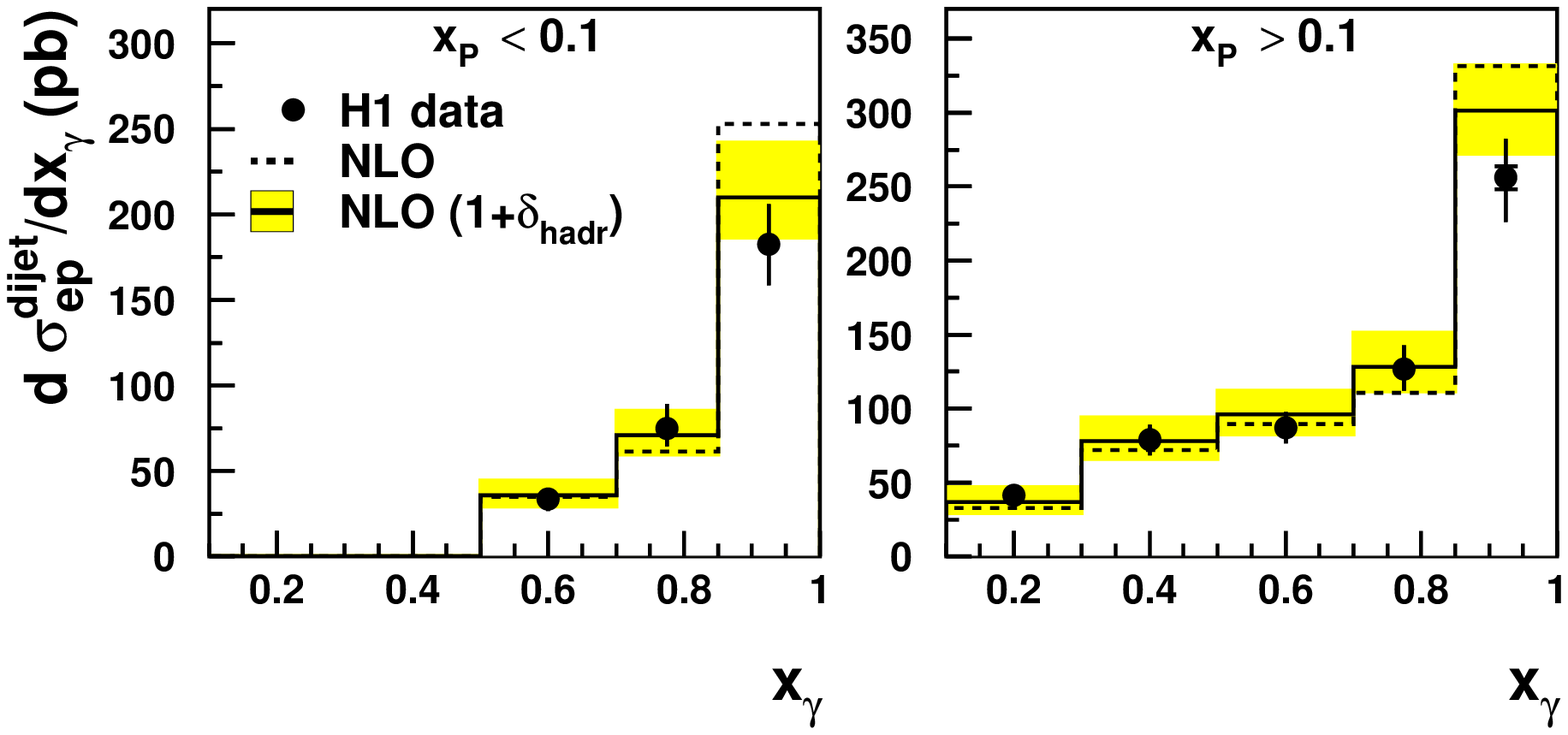} \\
      \includegraphics[width=\figwidth, clip=]{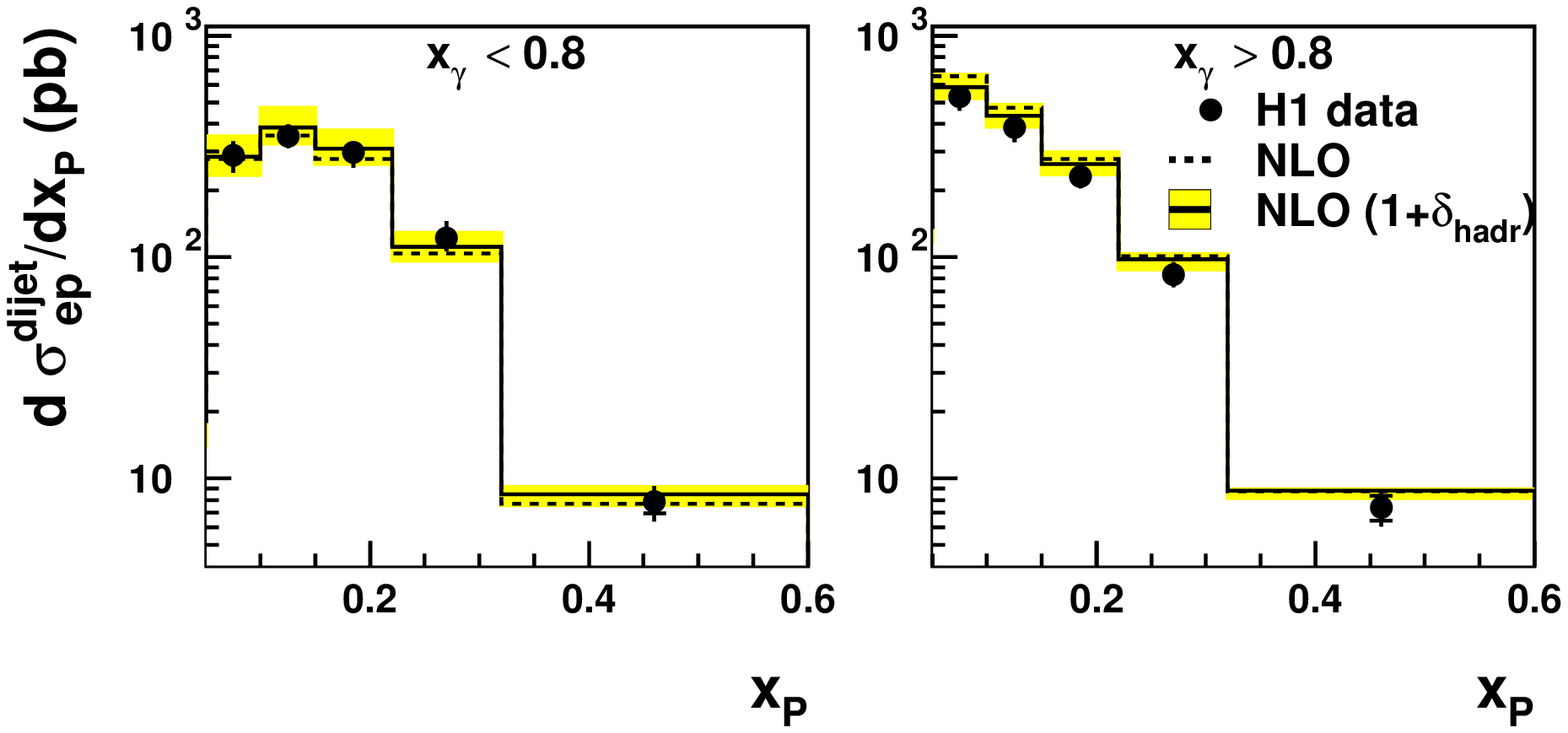}
    \end{tabular}
    \caption{\textit{%
        Dijet cross sections in photoproduction.
        }
      }
    \label{fig:h1-phpdijet}
  \end{center}
\end{figure}

Studies have been made by both H1~\cite{hep-ex-0201006} and
ZEUS~\cite{hep-ex-0112029} of dijets in photoproduction events. Leading order
calculations did not describe the rates particularly well. The cross section
has been measured as a function of \Et, $x_{\gamma}$ and $x_{p}$, where
$x_{\gamma}$ ($x_{p}$) is the fraction of the photon (proton) momentum that
participates in the hard scattering. The H1 results are shown in
fig.~\ref{fig:h1-phpdijet}.
The need for the NLO calculations is clear, particularly at high
$x_{\gamma}$. 

%
%
\section{The Highest \Qsq{} Range}
\label{sec:hiq2}

\begin{figure}[htbp]
  \begin{center}
    \begin{tabular}{ll}
      \includegraphics[width=7cm, bb=25 30 520 540, clip=]{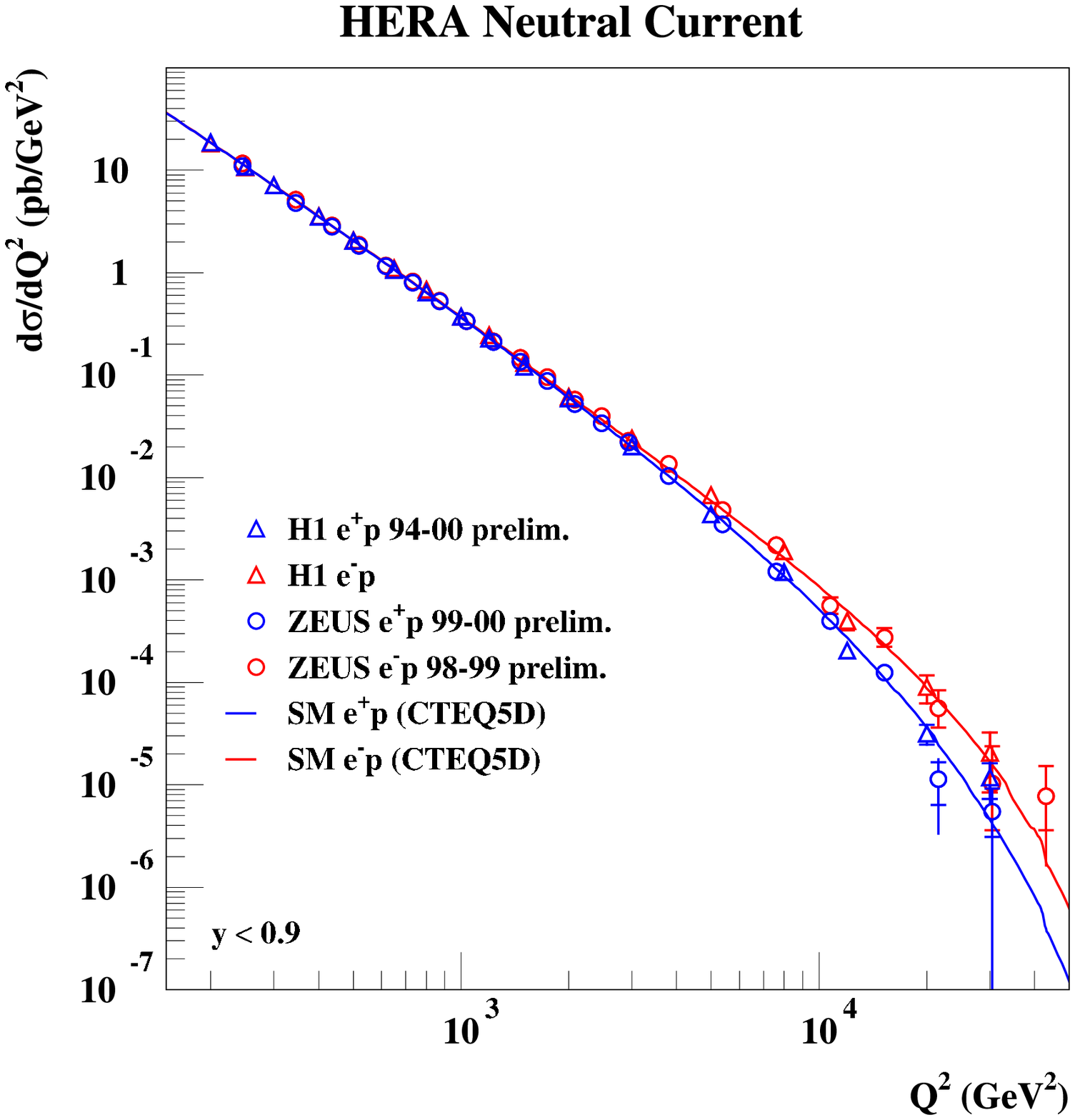} &
      \includegraphics[width=7cm, bb=25 30 520 540, clip=]{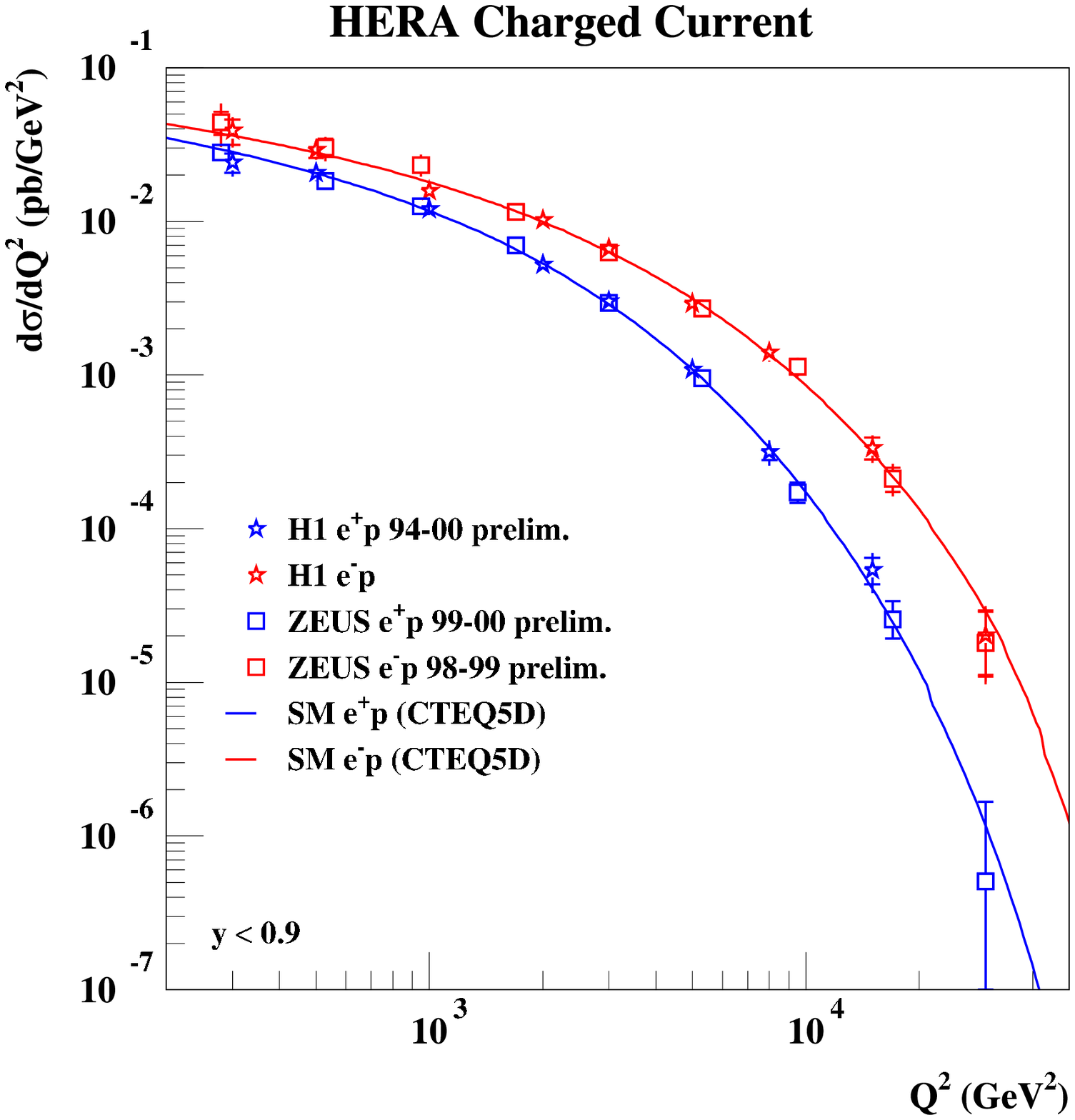}
    \end{tabular}
    \caption{\textit{%
        Neutral and charged current cross sections as measured by 
        H1 and ZEUS.
        }
      }
    \label{fig:nccc}
  \end{center}
\end{figure}

At the highest values of \Qsq, both H1 and ZEUS saw an excess in the cross
section with the data collected up to 1995. Newer data did not support this
excess. The cross sections for both neutral and charged current scattering now
include the data taken up to the end of 
2000~\cite{misc:eps01:630,*misc:eps01:631,*misc:eps01:787}. In both
neutral and charged current scattering very good agreement with the standard
model predictions using the CTEQ5 description of the PDFs is seen
(fig.~\ref{fig:nccc}).

%
%
\section{Heavy Quark Production}
\label{sec:hf}

Charm production at HERA has been studied for quite a while and there are
already a large number of publications. Data on bottom production has only
become available more recently with the steady increase in HERA luminosity.

As the dominant heavy flavour production mechanism is boson-gluon fusion,
measurements of charm production give a direct handle on the gluon content of
the proton. The results are consistent with those obtained from the fits
described in Section~\ref{sec:pdf}, but are not yet at the same level of
statistical precision.

\begin{figure}[htbp]
  \begin{center}
    \begin{tabular}{ll}
      \includegraphics[width=7cm]{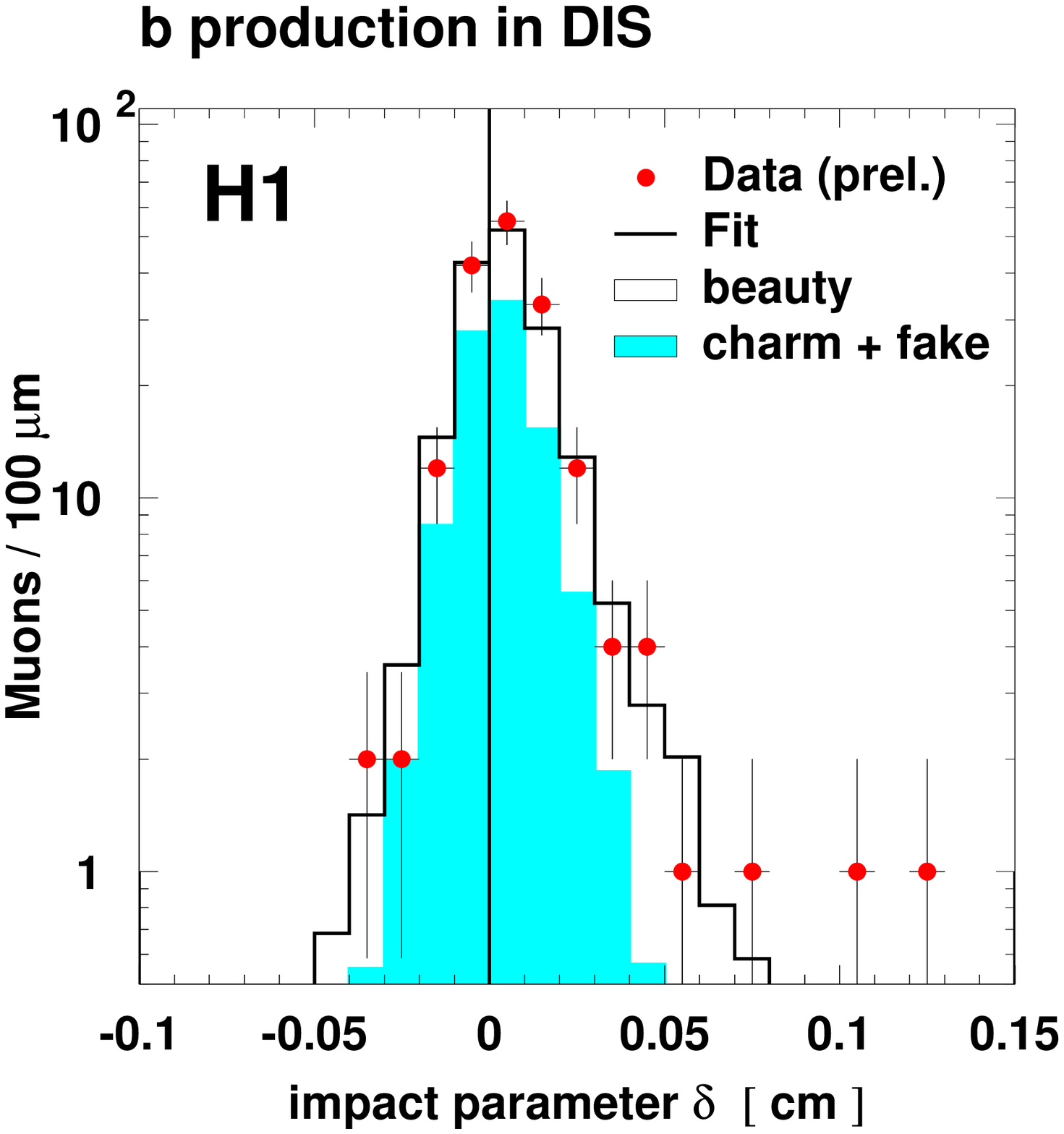} &
      \includegraphics[width=7cm]{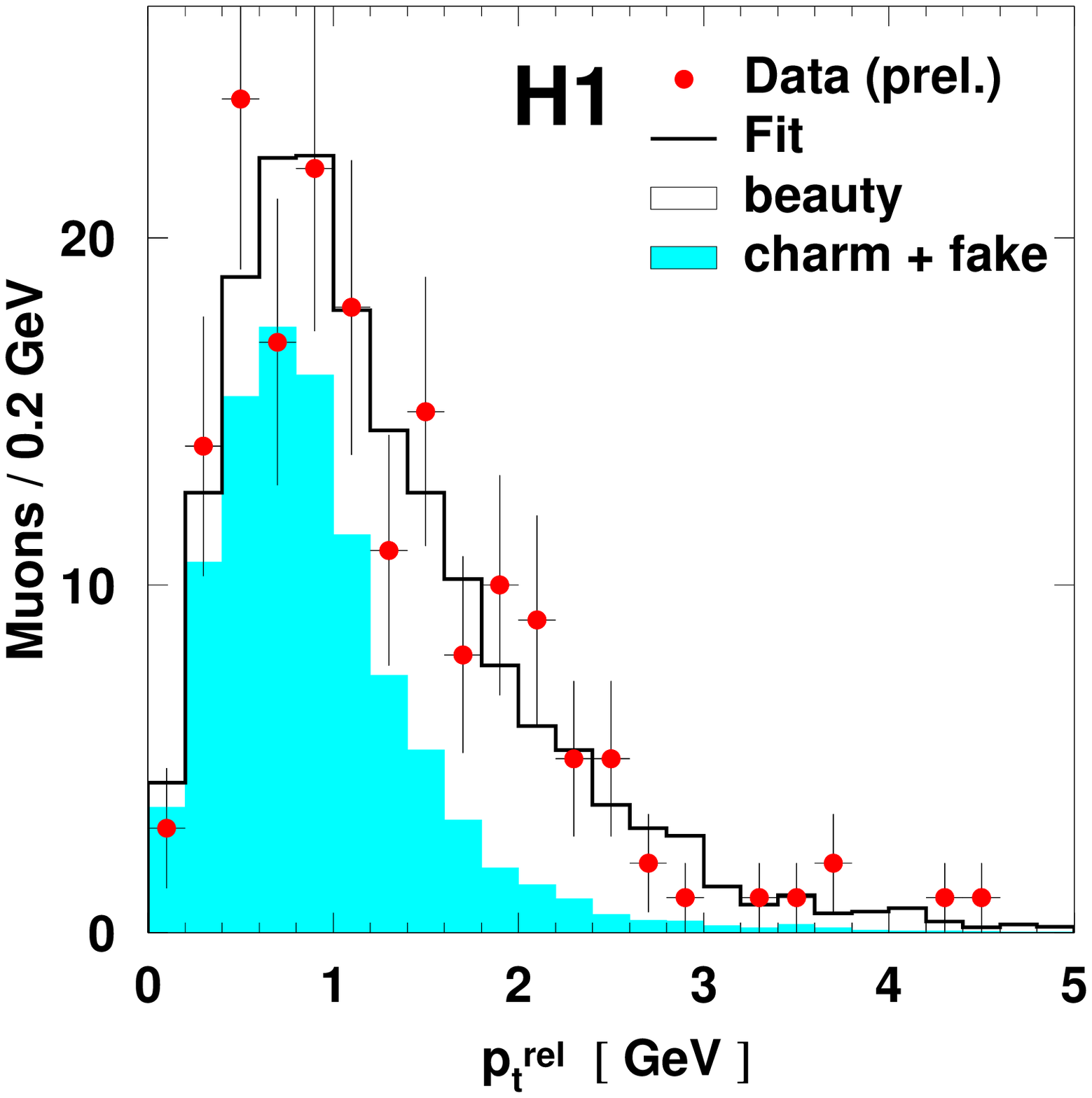}
    \end{tabular}
    \caption{\textit{%
        b quark signals as seen by H1 in DIS.
        }
      }
    \label{fig:h1-bsig}
  \end{center}
\end{figure}

\begin{figure}[htbp]
  \begin{center}
    \includegraphics[width=7cm]{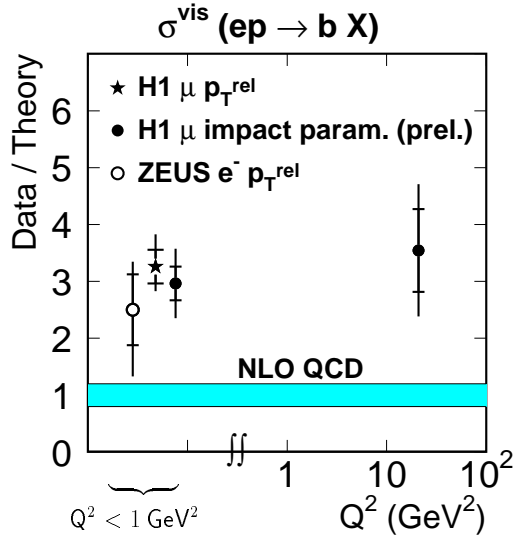}
    \caption{\textit{%
        H1 and ZEUS b quark production cross sections as a function of
        \Qsq.
        }
      }
    \label{fig:bxsec}
  \end{center}
\end{figure}

For both charm and bottom quark production, the mass of the quark provides the
hard scale necessary for perturbative QCD calculations. One would also expect
that the calculations for $b$ quark production should be more reliable than
those for $c$ quarks. H1 measured the $b$ quark production cross section for
the first time a couple of years ago and found it to be significantly higher
than the expectation~\cite{pl:b467:156}. They have now extended their analysis
to DIS~\cite{h1:eps01:807}. An excess at about the same level is seen. They
use a combination of a muon tag and impact parameter measurements and achieve
a signal to background ratio of about 1:1. In fig.~\ref{fig:h1-bsig} one can
see a clear $b$ signal at high values of impact parameter and transverse
momentum. The measured cross sections are shown in fig.~\ref{fig:bxsec}.
The recent ZEUS measurements~\cite{zeus:eps01:496} are consistent with those
from H1. ZEUS has also looked at Monte Carlos that include $b$ excitation.
While the predicted cross-sections are higher, they still lie below the data.

%
%
\section{Conclusions}
\label{sec:conclusions}

The HERA data provide many precise tests of perturbative QCD. NLO predictions
are necessary for many processes and more and more such predictions are now
available. In general they show an impressive level of agreement over a huge
kinematic range. However, in order to make further progress more input is
needed on what are appropriate uncertainties to use for the renormalisation
and factorisation scales. Some NNLO calculations have recently been completed
and more would be very welcome.

A number of accurate measurements of \alphas{} have been made by both
experiments that agree well with each other and with the world average. Many
different aspects of jet production have been measured. If one restricts the
kinematic range to regions where the NLO to LO corrections factors are not too
large, or to regions where the hadron to parton corrections are small, very
good agreement with the NLO predictions are seen.

While one would naively expect heavy flavour production to agree well with
perturbative QCD calculations, this is not the case. Charm cross sections are
in poor agreement with expectations, particularly in the forward direction;
bottom production is about a factor of two to three higher than the
predictions, both in photoproduction and DIS. Statistics are still very
limited here and the results will benefit from the HERA II luminosity upgrade
and the use of microvertex detectors, which have now been installed in both
experiments. We are looking forward to the 1\xifb{} that should be provided
for each experiment by 2006, and to the new possibilities offered by the
longitudinal beam polarisation.

%
%
%
\section{Acknowledgements}
I would like to thank the organisers for making this an interesting and
stimulating conference.  Several members of the ZEUS and H1 collaborations
provided valuable input during the preparation of this talk; in particular
Malcolm Derrick, Peter Schleper, Oscar Gonzalez, Claudia Glasman and Rik
Yoshida.


%
%
\catcode`\@=11 
\renewcommand{\@biblabel}[1]{#1.}
\catcode`\@=12 
\providecommand{\etal}{et al.\xspace}
\providecommand{\coll}{Coll.\xspace}
\catcode`\@=11
\def\@bibitem#1{%
\ifmc@bstsupport
  \mc@iftail{#1}%
    {;\newline\ignorespaces}%
    {\ifmc@first\else.\fi\orig@bibitem{#1}}
  \mc@firstfalse
\else
  \mc@iftail{#1}%
    {\ignorespaces}%
    {\orig@bibitem{#1}}%
\fi}%
\catcode`\@=12


\begin{mcbibliography}{10}

\bibitem{jetp:46:641}
Yu.L.~Dokshitzer,
\newblock Sov.\ Phys.\ JETP{} {\bf 46},~641~(1977)\relax
\relax
\bibitem{np:b126:298}
G.~Altarelli and G.~Parisi,
\newblock Nucl.\ Phys.{} {\bf B~126},~298~(1977)\relax
\relax
\bibitem{sovjnp:15:438}
V.N.~Gribov and L.N.~Lipatov,
\newblock Sov.\ J.\ Nucl.\ Phys.{} {\bf 15},~438~(1972)\relax
\relax
\bibitem{sovjnp:20:94}
L.N.~Lipatov,
\newblock Sov.\ J.\ Nucl.\ Phys.{} {\bf 20},~94~(1975)\relax
\relax
\bibitem{misc:eps01:628}
ZEUS~\coll,
\newblock {\em The {ZEUS} {NLO} {QCD} Fit to Determine Parton Distribution
  Functions and $\alpha_s$}, 2001.
\newblock Paper 628 submitted to the International Europhysics Conference on
  High Energy Physics, Budapest, Hungary, July 12-18, 2001\relax
\relax
\bibitem{h1:eps01:787}
H1~\coll,
\newblock {\em Inclusive measurement of deep inelastic scattering at high
  {$Q^{2}$} in {$ep$} collisions at {HERA}}.
\newblock Paper 787 submitted to the International Europhysics Conference on
  High Energy Physics 01, Budapest, Hungary, July 12-18 2001., 2001\relax
\relax
\bibitem{thesis:bornheim:1999}
A.~Bornheim,
\newblock {\em Messung der Protonstrukturfunktionen $F_2$ und $F_L$ in
  radiativer $ep$-Streuung mit dem ZEUS-Detektor}.
\newblock Ph.D.\ Thesis, Universit\"at Bonn, Report \mbox{BONN-IR-99-17},
  1999\relax
\relax
\bibitem{epj:c21:33}
H1 \coll, C.~Adloff \etal,
\newblock Eur.\ Phys.\ J.{} {\bf C~21},~33~(2001)\relax
\relax
\bibitem{epj:c24:33}
H1 \coll, C.~Adloff \etal,
\newblock Eur.\ Phys.\ J.{} {\bf C~24},~33~(2002)\relax
\relax
\bibitem{zeus:eps01:637}
ZEUS~\coll,
\newblock {\em Inclusive jet cross sections in neutral current deep inelastic
  scattering in the Breit frame and determination of {$\alpha_S$} at {HERA}}.
\newblock Paper 637 submitted to the International Europhysics Conference on
  High Energy Physics 01, Budapest, Hungary, July 12-18 2001., 2001\relax
\relax
\bibitem{epj:c19:289}
H1 \coll, C.~Adloff \etal,
\newblock Eur.\ Phys.\ J.{} {\bf C~19},~289~(2001)\relax
\relax
\bibitem{pl:b507:70}
ZEUS \coll, J.~Breitweg \etal,
\newblock Phys.\ Lett.{} {\bf B~507},~70~(2001)\relax
\relax
\bibitem{zeus:eps01:641}
ZEUS~\coll,
\newblock {\em Measurements of jet substructure in neutral current deep
  inelastic scattering and determination of {$\alpha_S$} at {HERA}}.
\newblock Paper 641 submitted to the International Europhysics Conference on
  High Energy Physics 01, Budapest, Hungary, July 12-18 2001., 2001\relax
\relax
\bibitem{pl:b515:17}
H1 \coll, C.~Adloff \etal,
\newblock Phys.\ Lett.{} {\bf B~515},~17~(2001)\relax
\relax
\bibitem{hep-ex-0201006}
H1 \coll, C.~Adloff \etal,
\newblock Preprint \mbox{DESY-01-225} (\mbox{hep-ex/0201006}), 2002.
\newblock Subm.\ to Eur.~Phys.~J\relax
\relax
\bibitem{hep-ex-0112029}
ZEUS \coll, S.~Chekanov \etal,
\newblock Preprint \mbox{DESY-01-220} (\mbox{hep-ex/0112029}), DESY, 2001.
\newblock Subm.\ to Eur.~Phys.~J\relax
\relax
\bibitem{misc:eps01:630}
ZEUS~\coll,
\newblock {\em Measurement of High-{$Q^2$} Neutral Current Cross Sections in
  $e^+p$ Deep Inelastic Scattering at {HERA}}.
\newblock Paper 630 submitted to the International Europhysics Conference on
  High Energy Physics, Budapest, Hungary, July 12-18, 2001, 2001\relax
\relax
\bibitem{misc:eps01:631}
ZEUS~\coll,
\newblock {\em Measurement of High-{$Q^2$} Charged Current Cross Sections in
  $e^+p$ Deep Inelastic Scattering at {HERA}}.
\newblock Paper 631 submitted to the International Europhysics Conference on
  High Energy Physics, Budapest, Hungary, July 12-18, 2001, 2001\relax
\relax
\bibitem{misc:eps01:787}
H1~\coll,
\newblock {\em Inclusive Measurement of Deep Inelastic Scattering at high $Q^2$
  in ep Collisions at {HERA}}.
\newblock Papers 787, 803 submitted to the International Europhysics Conference
  on High Energy Physics, Budapest, Hungary, July 12-18, 2001, 2001\relax
\relax
\bibitem{pl:b467:156}
H1 \coll, C.~Adloff \etal,
\newblock Phys.\ Lett.{} {\bf B~467},~156~(1999)\relax
\relax
\bibitem{h1:eps01:807}
H1~\coll,
\newblock {\em Beauty Production in Deep Inelastic {$ep$} Scattering}.
\newblock Paper 807 submitted to the International Europhysics Conference on
  High Energy Physics 01, Budapest, Hungary, July 12-18 2001., 2001\relax
\relax
\bibitem{zeus:eps01:496}
ZEUS~\coll,
\newblock {\em Beauty Photoproduction in the Muon Semi-Leptonic Decay Mode at
  {HERA}}.
\newblock Paper 496 submitted to the International Europhysics Conference on
  High Energy Physics 01, Budapest, Hungary, July 12-18 2001., 2001\relax
\relax
\end{mcbibliography}
\end{document}